\documentclass[aps,prl,reprint,groupedaddress,longbibliography]{revtex4-2}

\usepackage{siunitx}
\usepackage{graphicx}
\usepackage{placeins}
\usepackage{textcomp}
\usepackage{amssymb}
\usepackage{amsmath}
\usepackage{color}
\usepackage{natbib}
\usepackage{epstopdf}
\usepackage{makecell}

\newcommand{\textapprox}{\raisebox{0.5ex}{\texttildelow}}

\newcommand{\ket}[1] {|#1 \rangle}

\begin{document}

\title{Single-shot readout of multiple donor electron spins with a gate-based sensor}


\author{M.R. Hogg}
\altaffiliation{Present address: Department of Physics, The University of Basel, 4056 Basel, Switzerland}
\affiliation{Australian Research Council Centre of Excellence for Quantum Computation and Communication Technology, School of Physics, University of New South Wales, Sydney, NSW 2052, Australia}

\author{P. Pakkiam}
\altaffiliation{Present address: School of Mathematics and Physics, The University of Queensland, 4067 Brisbane, Australia}
\affiliation{Australian Research Council Centre of Excellence for Quantum Computation and Communication Technology, School of Physics, University of New South Wales, Sydney, NSW 2052, Australia}

\author{S.K. Gorman, A.V. Timofeev, Y. Chung, G. K. Gulati}
\affiliation{Australian Research Council Centre of Excellence for Quantum Computation and Communication Technology, School of Physics, University of New South Wales, Sydney, NSW 2052, Australia}

\author{M.G. House}
\altaffiliation{Present address: PsiQuantum Ltd, 94304 Palo Alto, United States}
\affiliation{Australian Research Council Centre of Excellence for Quantum Computation and Communication Technology, School of Physics, University of New South Wales, Sydney, NSW 2052, Australia}

\author{M.Y. Simmons}
\email[Corresponding author: ]{michelle.simmons@unsw.edu.au}
\affiliation{Australian Research Council Centre of Excellence for Quantum Computation and Communication Technology, School of Physics, University of New South Wales, Sydney, NSW 2052, Australia}



\begin{abstract}
Proposals for large-scale semiconductor spin-based quantum computers require high-fidelity single-shot qubit readout to perform error correction and read out qubit registers at the end of a computation. 
However, as devices scale to larger qubit numbers integrating readout sensors into densely packed qubit chips is a critical challenge.
Two promising approaches are minimising the footprint of the sensors, and extending the range of each sensor to read more qubits.
Here we show high-fidelity single-shot electron spin readout using a nanoscale single-lead quantum dot (SLQD) sensor that is both compact and capable of reading multiple qubits. 
Our gate-based SLQD sensor is deployed in an all-epitaxial silicon donor spin qubit device, and we demonstrate single-shot readout of three $^{31}$P donor quantum dot electron spins with a maximum fidelity of 95\%.
Importantly in our device the quantum dot confinement potentials are provided inherently by the donors, removing the need for additional metallic confinement gates and resulting in strong capacitive interactions between sensor and donor quantum dots. We observe a $1/d^{1.4}$ scaling of the capacitive coupling between sensor and $^{31}$P dots (where $d$ is the sensor-dot distance), compared to $1/d^{2.5-3.0}$ in gate-defined quantum dot devices. Due to the small qubit size and strong capacitive interactions in all-epitaxial donor devices, we estimate a single sensor can achieve single-shot readout of approximately 15 qubits in a linear array, compared to 3-4 qubits for a similar sensor in a gate-defined quantum dot device. Our results highlight the potential for spin qubit devices with significantly reduced sensor densities.
\end{abstract}


\maketitle

\section{Introduction}
\noindent
High-fidelity single-shot spin readout is an essential requirement for large-scale spin-based quantum computing proposals \cite{Hill2015, Veldhorst2017}.
In spin qubit devices readout has most commonly been performed by mapping the spin state onto a charge state, which can then be detected using a nearby charge sensor \cite{Elzerman2004, Morello2010}. The most sensitive charge sensors to date are radio-frequency single-electron transistors (rf-SETs) which have demonstrated integration times for a signal-to-noise (SNR) ratio of 1 of $\tau_{min} = 0.62$\,ns \cite{Keith2019} (we use $\tau_{min}$ to directly compare sensor performance between experiments, see Supplementary S1). However rf-SETs require at least three gate leads to operate, a significant space allocation on the qubit chip as devices scale to larger qubit numbers. In contrast gate-based (or dispersive) sensors integrate qubit readout capability into the existing leads used for qubit control, allowing spin measurements without additional proximal charge sensors \cite{Petersson2010, Colless2013, House2015, Gonzalez-Zalba2015}. 
Single-shot readout using dispersive sensors has recently been demonstrated \cite{Pakkiam2018, West2019, Urdampilleta2019, Zheng2019, Borjans2021}, however these results relied on Pauli blockade between two spins for readout, requiring an additional quantum dot (with associated control leads) for spin detection. 
Furthermore, in each case readout of only a single qubit was achieved. Dispersive sensors which probe inter-dot transitions (referred to here as ``direct dispersive'' sensors) can also only generate signal at an inter-dot charge transition, a small window in gate space. Experiments to date have thus often used a secondary multi-lead charge sensor for device characterisation \cite{Pakkiam2018, West2019}. 

The SLQD is a gate-based sensor that combines the benefits of rf-SETs and direct dispersive sensors \cite{House2016}. It requires only a single lead to operate, improving scalability prospects over the rf-SET, yet can also perform readout directly in the single-spin basis. 
The SLQD is sensitive over a wide range of device voltages, and has been used as a compact sensor for single-electron charge detection in quantum dot arrays \cite{Chanrion2020, Ansaloni2020, duan2020}, as well as time-averaged single-spin measurements \cite{Cirianotejel2020}. 

In this work we deploy a SLQD charge sensor in a device containing four $^{31}$P donor quantum dots. We use the SLQD to perform single-shot electron spin readout in the single-spin basis on three out of the four quantum dots, achieving a maximum readout fidelity of 95\%. The fourth quantum dot was found to have a tunnel rate exceeding the measurement bandwidth (300\,kHz) and could not be projectively measured. 
We achieve an amplitude signal to noise ratio of 9.6 at 15\,kHz measurement bandwidth, yielding $\tau_{min} = 720$\,ns. This compares favourably to previous dispersive readout measurements operating with off-chip resonators at 100-300\,MHz frequencies \cite{Pakkiam2018, West2019, Urdampilleta2019} (respectively $\tau_{min} = 19\mathrm{\mu s}$, $3000\mathrm{\mu s}$ and $8\mathrm{\mu s}$).

Importantly we find that for our donor device the detection range of the sensor is extended compared to gate-defined quantum dot architectures. Charge sensor range is determined by the capacitive coupling between the sensor and the readout target, which in our device follows a $1/d^{1.4\pm0.1}$ dependence as a function of sensor-qubit distance, $d$.
Previous reports in undoped planar SiGe devices observed a $1/d^3$ dependence \cite{Zajac2016, Neyens2019}, and in nanowire CMOS devices a $1/d^{2.5-2.8}$ dependence was estimated based on device simulations \cite{duan2020}. In free space the capacitive coupling between isolated charges scales as $1/d$, however this scaling is modified by the presence of surrounding gates and becomes $1/d^3$ for charges underneath a large metal plane \cite{Neyens2019}, as is the case for gate-defined devices operated in accumulation mode \cite{Borselli2015, Lim2009b}.
In atomically engineered donor devices the qubit trapping potential is defined naturally by the donor, eliminating the need for additional accumulation gates and resulting in extremely low gate densities. The reduced screening in atomic precision donor devices considerably extends charge sensor range. We estimate that our current sensor would be able to read out approximately 15 donor qubits in a linear array with high fidelity, compared to 3-4 for a sensor with similar performance in a gate-defined quantum dot device.

\section{Device and sensor operation}
\begin{figure}[t!]
 \includegraphics[width=0.5\textwidth]{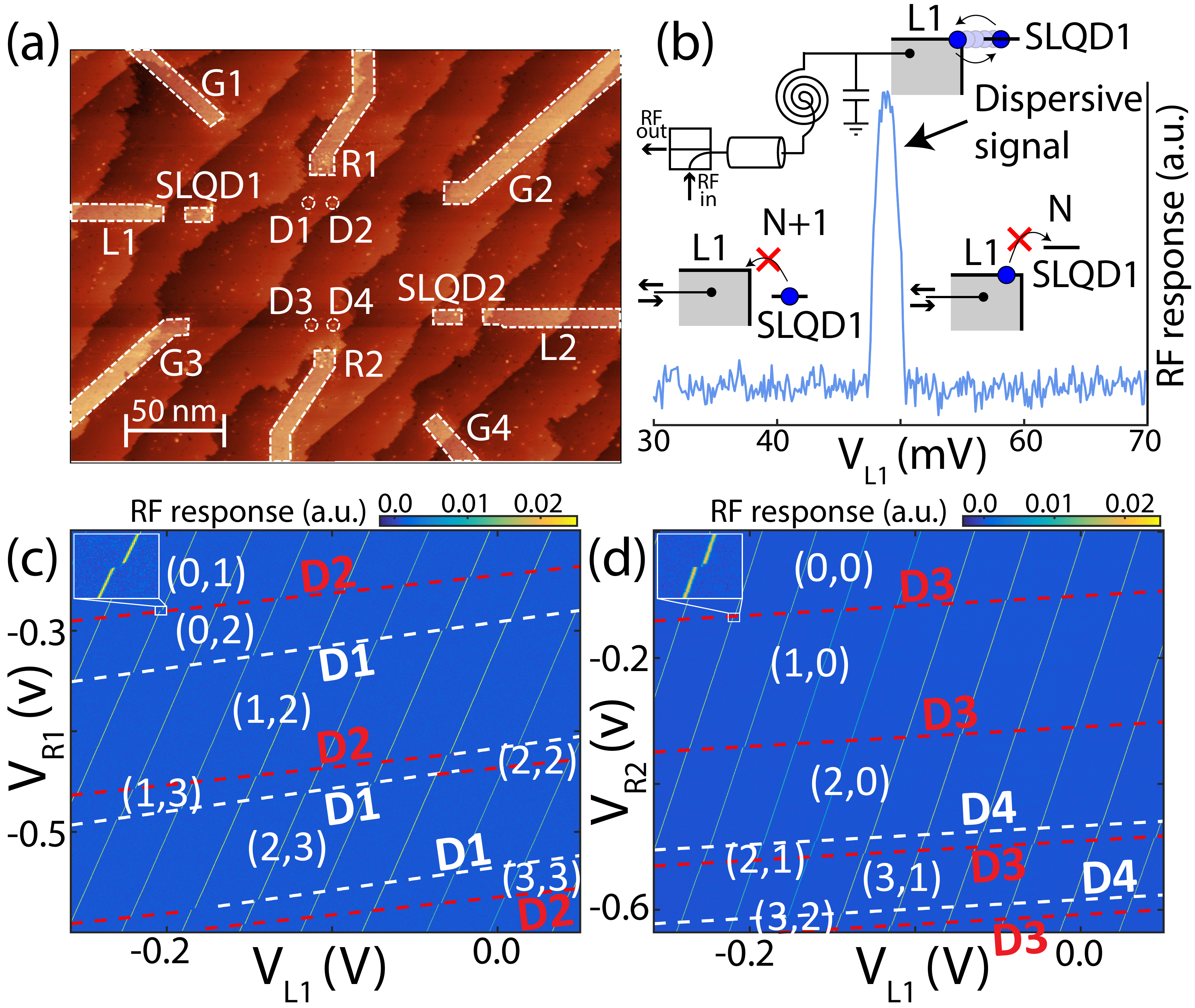}%
 \centering
 \caption{\textbf{Four donor-based quantum dot qubits with integrated SLQD charge sensors.} 
 	\textbf{(a)} STM image of device and gate labels. SLQD1 is used to sense the donor quantum dots (labelled D1, D2, D3 and D4). Based on STM images and electrically measured charging energies, we estimate 
 	the number of $^{31}$P donors in each to be 2 for D1, 3 for D2, 3 for D3 and 1 for D4.
 	Electrons are loaded onto the donors from R1 (D1, D2) or R2 (D3, D4) and onto SLQD1 from L1. There is a second SLQD in the device (SLQD2) that is not used in our experiments.
 	\textbf{(b)} Schematic of SLQD charge sensor operating principle. The sensor comprises a quantum dot (SLQD1) tunnel coupled to a single lead (L1) embedded in a resonant circuit formed by a NbTiN superconducting spiral inductor, measured using standard reflectometry techniques \cite{Petersson2010}. 
 	When an AC excitation is applied to L1,
 	cyclical single-electron tunneling generates quantum capacitance in the circuit when the Fermi energy in L1 is aligned with an available charge state of SLQD1, resulting in regular Coulomb-like peaks in the reflected signal.
 	 A change in the electrostatic environment causes the peaks to shift, allowing operation as a charge sensor.
 	\textbf{(c)} Charge stability diagram for D1 and D2 sweeping $V_\mathrm{L1}$ and $V_\mathrm{R1}$. The periodic diagonal lines in the data are SLQD1 transition lines, with donor charge transitions from D1 and D2 being observed as breaks in the SLQD lines (example break shown in inset). The overlaid white dotted lines indicate D1 charge transitions and the red dotted lines indicate D2 charge transitions, as labelled.
 	\textbf{(d)} Similar charge stability diagram for D3 and D4 sweeping $V_\mathrm{L1}$ and $V_\mathrm{R2}$. The red dotted lines now indicate D3 charge transitions, and the white dotted lines D4 charge transitions.}
 \label{fig:STM_image}
\end{figure}

The four $^{31}$P quantum dot device is fabricated on a silicon substrate using atomic precision hydrogen resist lithography \cite{Simmons2005, Fuechsle2012}. Figure \ref{fig:STM_image}\,(a) shows an STM image of the device, with sites D1, D2, D3 and D4 indicating regions where phosphorus donors are incorporated to host electron spin qubits \cite{He2019}. R1 and R2 serve as electron reservoirs for (D1, D2) and (D3, D4) respectively, as well as providing electrostatic tuning of the donor potentials.
G1 and G2 are electrostatic gates used for readout pulse sequences on D1 and D2, and G3 and G4 serve the same purpose for D3 and D4. SLQD1 and SLQD2 are charge sensors for determining the electron occupation of the donor dots, and to perform single-shot spin readout via spin to charge conversion.

A schematic of the operating principle of the SLQD sensor is shown in Fig. \ref{fig:STM_image}\,(b). The sensor detects the extra quantum capacitance \cite{Petersson2010, House2015, Gonzalez-Zalba2015} added to the circuit when the Fermi energy of L1 aligns with an available charge state on SLQD1, resulting in Coulomb-like peaks that can be used for charge sensing.
We note that in this work SLQD1 is used for all measurements.
Figures \ref{fig:STM_image}\,(c),(d) show charge stability diagrams for the top (D1, D2) and bottom (D3, D4) pairs of donor quantum dots, respectively. 
The periodic diagonal lines in the data are Coulomb-like peaks from SLQD1, where donor charge transitions are observed as breaks in these lines aligned along the labelled dotted lines overlaying the data. 
We identify transitions from all four patterned donor quantum dots from the distinct transition line slopes, shown in Fig. \ref{fig:STM_image}\,(c) (D1 white dashed lines, D2 red dashed lines) and Fig. \ref{fig:STM_image}\,(d) (D3 red dashed lines, D4 white dashed lines).
Electron numbers are assigned by fully depleting the donors of electrons, then adding an electron each time a donor transition line is crossed. 
We note that sweeping R1 (R2) to negative voltages adds electrons to D1 and D2 (D3 and D4), in contrast to sweeping gates that are only capacitively coupled (and not tunnel-coupled), for which negative voltages typically remove electrons. 
These plots demonstrate the ability of SLQD1 to characterise the charge occupation of all four quantum dots. 

\section{Single-shot spin readout}
\begin{figure}[t!]
	\includegraphics[width=0.48\textwidth]{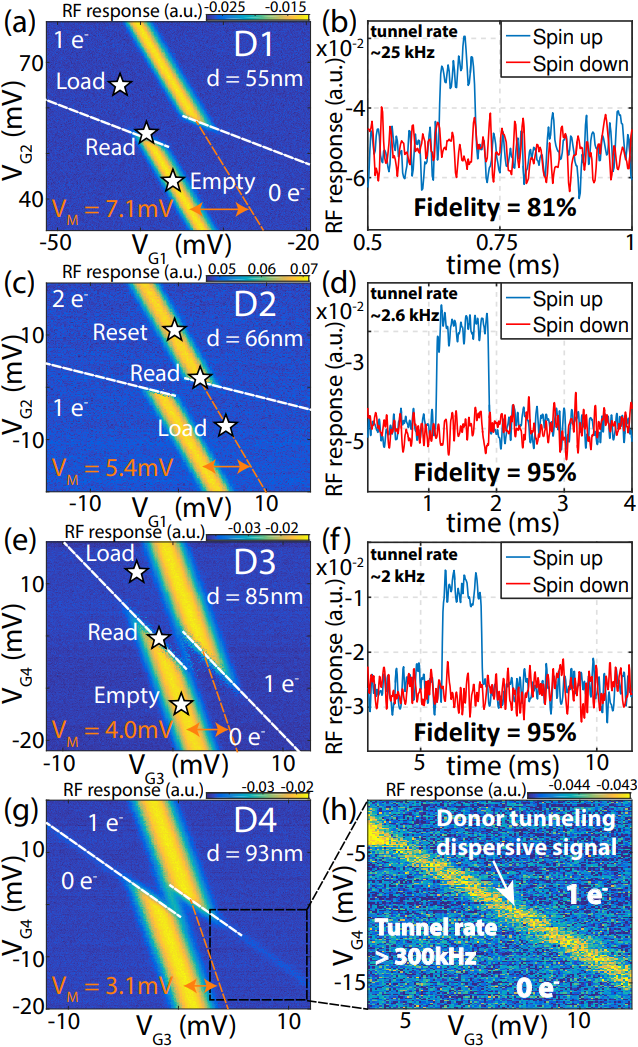}%
	\caption{\textbf{Single-shot electron spin readout on D1, D2 and D3 donor quantum dots.} 
		\textbf{(a)} Gate-gate map showing break in SLQD1 transition line when the first electron is loaded onto D1. The approximate pulse positions for the readout sequence are indicated by the white stars. The read level is calibrated so that a spin up electron can tunnel from D1 to R1, but a spin down cannot.
		\textbf{(b)} Example single-shot readout traces showing spin up and spin down signals. Due to the fast tunnel rate at this transition, some of the rapid spin up traces are not registered, leading to an average fidelity of 81\%. 
		\textbf{(c)} Similar gate-gate map for D2. Readout for D2 was performed at the second electron transition via the D$^-$ charge state \cite{Watson2015}. Here we chose the read position on the opposite side of the SLQD charge break to make spin-up the high sensor level for consistency with D1 and D3. 
		\textbf{(d)} Example single-shot readout traces for D2. 
		\textbf{(e)} Gate-gate map for D3. 
		\textbf{(f)} Example single-shot readout traces for D3. 
		\textbf{(g)} Gate-gate map for D4. For this donor quantum dot, the tunnel rate between the donor and reservoir was too fast to perform single-shot spin readout. In fact, a faint signal can be observed due to cyclical driving of an electron between the D4 and R2. This implies that the tunnel rate is non-negligible compared to the RF reflectrometry frequency (130\,MHz). 
		\textbf{(h)} Zoom in on region from (g), highlighting the dispersive signal generated by cyclical tunneling of the D4 electron.}
	\label{fig:spin_readout}
\end{figure}

We next use SLQD1 to perform single-shot electron spin readout in the single-spin basis for D1, D2 and D3 at B=1.5\,T. We note that for D4 the tunnel rate to R2 exceeded our measurement bandwidth and readout could not be performed.
Figure \ref{fig:spin_readout}\,(a) shows a gate-gate map sweeping $V_\mathrm{G1}$ and $V_\mathrm{G2}$ over a single SLQD1 line intersecting with the first electron charge transition of D1. Adding an electron to D1 shifts the SLQD1 peak position by $V_\mathrm{M}=7.1$\,mV (where $V_\mathrm{M}$ is the mutual charging voltage in terms of $V_\mathrm{G1}$). We perform spin readout with a three-level pulse sequence consisting of a load phase to initialise a random electron spin state, followed by a read phase where the spin is projectively measured, and an empty phase to eject the electron before the next pulse repetition \cite{Morello2010}.
During the read phase a spin up state will tunnel to R1 followed by a spin down tunnelling back to D1, generating a characteristic ``blip'' in the charge sensor response which is absent for a spin down state. Figure \ref{fig:spin_readout}\,(b) shows example spin up and spin down traces for D1, demonstrating single-shot readout. We calculate the spin readout fidelity by taking 5000 individual single-shot traces and find $F_M=81\%$ (see Supplementary S3 for full details of the fidelity analysis). The fidelity is limited by our measurement bandwidth (set to \textapprox80\,kHz), which was not high enough to capture the fastest tunnelling events. Increasing the measurement bandwidth permits faster events at the cost of reduced SNR (which also decreases the fidelity \cite{Keith2019a}), and we found 80\,kHz to give the maximum fidelity of 81\%.

Figure \ref{fig:spin_readout}\,(c) shows a gate-gate map at the second electron of D2. Here we performed $D^-$ readout \cite{Watson2015} due to the favourable electron tunnel rate at the second electron transition (\textapprox2.6\,kHz). Example single-shot traces are shown in Fig. \ref{fig:spin_readout}\,(d), and taking 5000 individual traces we find a fidelity of $F_M=95\%$ for D2. This fidelity is again limited in part by our finite measurement bandwidth (here 15\,kHz) filtering the fastest tunnelling events, as well as a relatively high electron temperature (\textapprox280\,mK). 
We note the electron temperature was not raised by the reflectometry signal applied to SLQD1 until significantly higher powers than used in our experiments.
Figure \ref{fig:spin_readout}\,(e) shows a gate-gate map at the first electron of D3. In this case electrons tunnel between D3 and R2 (rather than R1 as in the previous cases) during spin readout. Figure \ref{fig:spin_readout}\,(f) shows corresponding example single-shot traces for D3, and here we find a fidelity of $F_M=95\%$, which is limited by the same factors as D2.

Figure \ref{fig:spin_readout}(g) shows a gate-gate map at the first electron transition of D4. Here we found that the D4-R2 tunnel rate exceeded our measurement bandwidth (\textapprox300\,kHz, limited by the resonator bandwidth), and we were not able to perform readout on this donor. In fact, a faint dispersive signal can be observed due to cyclical tunneling of an electron between D4 and R2 (shown in Fig. \ref{fig:spin_readout}\,(h)),
implying that this tunnel rate is a non-negligible fraction of the \textapprox130\,MHz reflectometry drive signal.
Table \ref{tab:donor_properties} summarises our spin readout results, with a maximum fidelity of 95\%. By lowering the electron temperature to $<160$\,mK, using a higher frequency resonant circuit and lower noise first-stage amplifier, readout of D1, D2 and D3 with $>$99\% fidelity is achievable \cite{Keith2019a, Watson2015, Schaal2020}.
\FloatBarrier
\onecolumngrid

\begin{table*}[hbtp]
	\centering
	\caption{\textbf{Summary of single-shot readout results for four donor dot device.}}
	\begin{tabular}{c|c|c|c|c|c|c|c|c|c}
		\hline
		\hline
		\rule{0pt}{2ex}
		& \makecell{Donor\\number} & \makecell{Readout\\transition\\electron \#}  & \makecell{Tunnel\\rate\\(kHz)} & \makecell{Distance\\to sensor\\(nm)} & \makecell{Sensor\\shift\\(mV)}  & \makecell{Readout\\fidelity\\(\%)} & \makecell{Measurement\\bandwidth\\(kHz)} & \makecell{Amplitude\\SNR} & $\tau_{min}$\,(ns) \\
		
		\hline
		D1    & 2P     & 0$\leftrightarrow$1  & \textapprox25   & 55    & 7.1   & 81 & 80 & 4.0 & 780\\
		D2    & 3P     & 1$\leftrightarrow$2  & \textapprox2.6  & 66    & 5.4   & 95 & 15 & 9.6 & 720\\
		D3    & 3P     & 0$\leftrightarrow$1  & \textapprox2.0  & 85    & 4.0   & 95 & 15 & 8.1 & 1020\\
		D4    & 1P     & 0$\leftrightarrow$1  & $>$300  		& 93    & 3.1   & N/A & N/A & N/A & N/A\\
		\hline
		\hline
	\end{tabular}%
	\label{tab:donor_properties}%
\end{table*}

\twocolumngrid

\section{Long-range charge sensing}
\begin{figure}[b!]
	\includegraphics[width=0.5\textwidth]{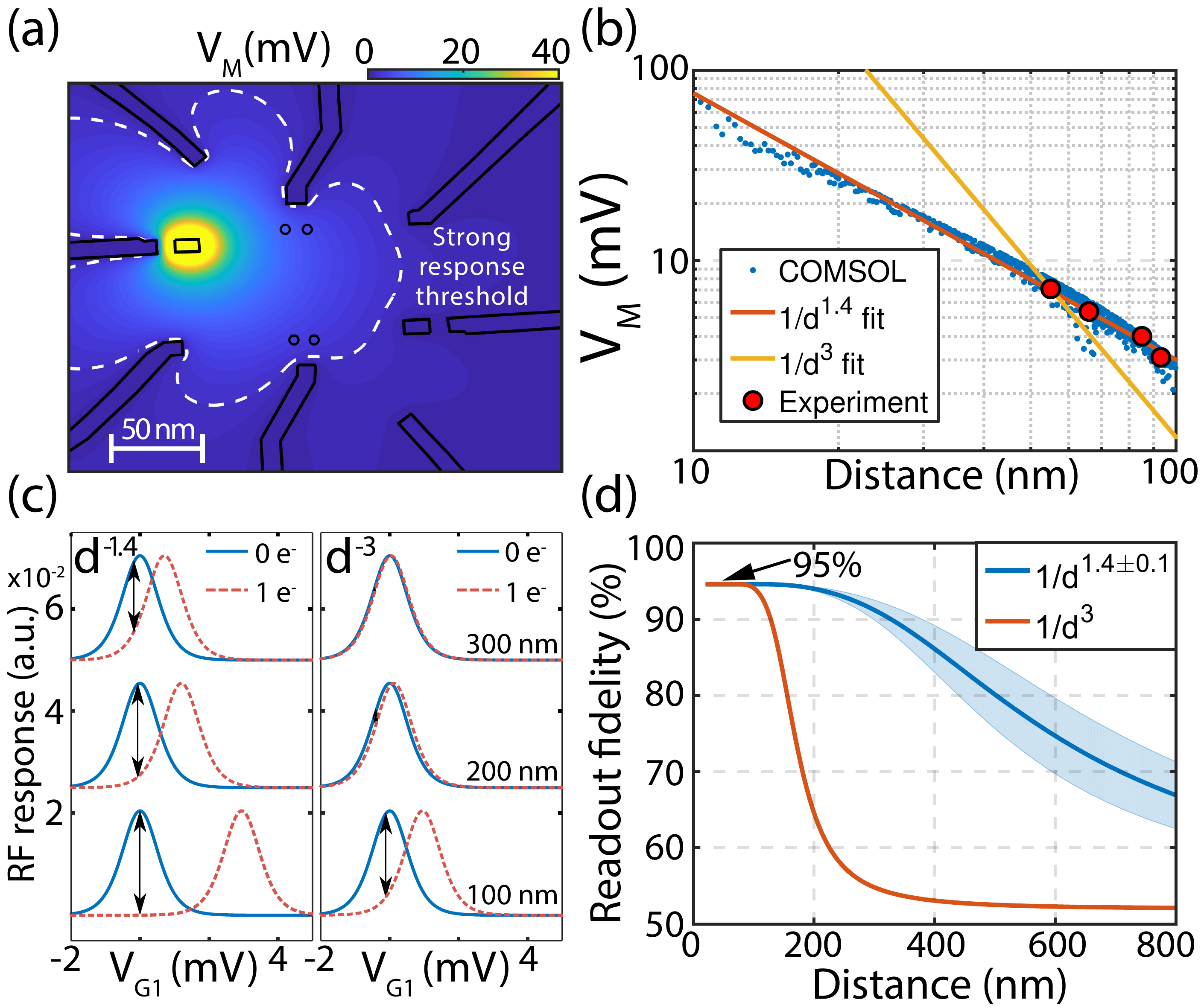}%
	\caption{\textbf{Importance of the sensor shift $V_\mathrm{M}$ for long-range qubit readout.}
	\textbf{(a)} COMSOL simulation of $V_\mathrm{M}$ for our device geometry. The dashed contour line indicates the estimated strong response regime boundary.
	\textbf{(b)} Fit to COMSOL simulated $V_\mathrm{M}$ values as a function of distance $d$ from SLQD1. The simulated values (blue dots) follow a $1/d^{1.4\pm0.1}$ dependence (orange line), which match the experimental values (red circles). The yellow line is a $1/d^3$ fit (as observed in planar gate-defined devices) to the experimental data, which is inconsistent with both the simulated and measured values.
	\textbf{(c)} Comparison of readout contrast possible for 100, 200 and 300\,nm sensor-qubit distance for $1/d^{1.4}$ scaling (left panel) and $1/d^3$ scaling (right panel). At 300\,nm the sensor peaks for two qubit charge states are almost overlapping in the $1/d^3$ case, whereas significant discrimination is still possible for $1/d^{1.4}$.
	\textbf{(d)} Estimated single-shot readout fidelity as a function of qubit distance to SLQD1. The input parameters to the fidelity calculation (tunnel rates, $T_1$ relaxation time, noise level) are taken from the experimentally measured values for D3. The blue curve is for $1/d^{1.4\pm0.1}$, with the shaded region outlining the uncertainty bounds from the fit in Fig. \ref{fig:distance_sensing}\,(b). The orange curve is for $1/d^3$, and exhibits a much faster rolloff in readout fidelity. Over 90\% fidelity is possible out to 300\,nm for $1/d^{1.4}$, compared to 130\,nm for $1/d^3$.}
	\label{fig:distance_sensing}
\end{figure}

We next investigate the scaling of the SLQD1 sensor shift $V_\mathrm{M}$ as a function of the sensor-qubit distance $d$. We used the finite element package COMSOL multiphysics to simulate $V_\mathrm{M}$ for the device in Fig. \ref{fig:STM_image}\,(a) (Supplementary S4). Figure \ref{fig:distance_sensing}\,(a) plots the simulated $V_\mathrm{M}$ profile, with the colourscale showing the expected $V_\mathrm{M}$ for a qubit at the corresponding position.
$V_\mathrm{M}$ determines the SLQD1 peak shift due to a charging event on the target qubit, and is proportional to the capacitive coupling between sensor and qubit. The white dashed contour line in Fig. \ref{fig:distance_sensing}\,(a) indicates the region within which $V_\mathrm{M}$ is large enough to
shift the SLQD1 Coulomb peak from full signal (on top of the peak) to $<1$\% signal due to a charging event on the target qubit (the strong response threshold).
Any qubit located inside the footprint of this contour would generate full on-off switching of the sensor signal \cite{Borjans2021}, and could be measured without any loss of fidelity due to the distance from the sensor.

Figure \ref{fig:distance_sensing}\,(b) plots the simulated $V_\mathrm{M}$ as a function of $d$ (the distance from the center of SLQD1) for the device region containing the sensor and patterned qubits (Supplementary S4).
Fitting to the simulated $V_\mathrm{M}$ values we find $V_\mathrm{M} \propto d^{1.4\pm0.1}$, which is consistent with the measured data (shown by red circles). Both our experimental and simulated results are consistent with a $1/d^{1.4}$ scaling, in contrast to the $1/d^3$ scaling observed in planar gate-defined devices \cite{Zajac2016, Neyens2019}, indicated by the yellow curve in Fig. \ref{fig:distance_sensing}\,(b).
Figure \ref{fig:distance_sensing}\,(c) highlights the impact this difference in scaling has for long-range charge sensing. The left panel uses the $1/d^{1.4}$ fit from Fig. \ref{fig:distance_sensing}\,(b) to estimate $V_\mathrm{M}$ at distances of 100, 200 and 300\,nm from the sensor. The solid blue and red dotted lines show the expected position of the SLQD Coulomb peak for 0 electrons and 1 electron on the target qubit at the specified distance. The shape of the sensor Coulomb peak is determined by fitting the experimental data. The vertical black arrows indicate the maximum sensor contrast for detecting the electron charge on the target. The right panel uses the $1/d^3$ fit from Fig. \ref{fig:distance_sensing}\,(b) to estimate $V_\mathrm{M}$ for 100, 200 and 300\,nm from the sensor. In this case $V_\mathrm{M}$ decreases more rapidly with distance, and the contrast is minimal at 300\,nm.

To highlight the importance of this roll-off for scalable quantum computing, we calculate the expected single-shot readout fidelity as a function of $d$ for a qubit with the same spin-dependent tunnel rates and T$_1$ time as D3 in our device. We extrapolate the fit in Fig. \ref{fig:distance_sensing}\,(b) to estimate $V_\mathrm{M}$ as a function of $d$, and use this to calculate the expected signal contrast. We can then calculate the SNR using the noise measured in our experiment and input this to our fidelity calculation (Supplementary S5). Figure \ref{fig:distance_sensing}\,(d) shows the result for both $1/d^{1.4\pm0.1}$ (blue curve) and $1/d^3$ (orange curve).
As expected the $1/d^3$ curve exhibits a faster roll-off in readout fidelity. For small $d$ both red and blue curves saturate to 95\% fidelity as observed in our direct experimental measurements of D3. We find that a qubit with the same properties as D3 could be measured with over 90\% spin readout fidelity up to 300\,nm from the sensor, compared to 130\,nm for the $1/d^3$ scaling.
We predict that the estimated 90\% readout fidelity at 300\,nm for our current sensor could be increased to $>$99\% by reducing the electron temperature \cite{Watson2015}, operating the SLQD at higher frequencies and using a lower noise first-stage amplifier \cite{Schaal2020}.

With a 300\,nm sensor range, the small size of donor qubits means that a single sensor has the potential to serve a large number of qubits. A leading proposal for a donor-based surface code architecture uses a donor separation distance of 30\,nm \cite{Hill2015}. Linear arrays are the current experimental state of the art in semiconductor spin qubits; for a linear array of donor qubits with an inter-qubit separation of 30\,nm, our results suggest that we would be able to read out $\lesssim$20 donor qubits with high fidelity with SLQD1 placed in the center of the array. To verify the predicted qubit number that could be sensed we directly simulated a realistic linear array of donor qubits (see Supplementary S6), from which we found that sensing 14-16 qubits is feasible using our current SLQD sensor.
For the geometry and scaling expected in an accumulation-mode gate-defined device ($1/d^3$, assuming a qubit separation of 80\,nm and sensor range of 130\,nm), the equivalent number is $\lesssim$4. This is consistent with current experimental demonstrations having 3-4 qubits per sensor \cite{Zajac2016, Volk}. Table \ref{tab:literature_comparison} compares our result to recently reported single-lead sensing literature, demonstrating the significant improvement in the number of possible qubits per sensor in a linear array (see Supplementary S5 for further details).

Given the extremely long $T_1$ relaxation times in donor qubits (we measured \textapprox9\,s for D3 at 2.5 Tesla, and up to 30\,s has been demonstrated \cite{Watson2017a}) and recent advances navigating charge states in large qubit arrays \cite{Volk, Mills2019}, 15 qubits similar to D3 could be measured sequentially by the same sensor within the $T_1$ relaxation time. We note that reducing the electron temperature would narrow the SLQD Coulomb peaks, increasing the charge detection signal for small $V_\mathrm{M}$ shifts and further extending the sensor range (and thus number of qubits measured) beyond our current estimate. We also note that donor qubits can be placed significantly closer than 30\,nm using atomic precision lithography. For an architecture of tunnel coupled donors (typically separated by 10-15\,nm), a single optimised SLQD sensor could conceivably serve a linear array of 50-100 donor qubits.

\FloatBarrier

\section{Conclusion}
In conclusion, we have demonstrated single-shot electron spin readout in the single-spin basis using a SLQD charge sensor. In a device with four $^{31}$P donor quantum dots we performed single-shot readout on three of the four quantum dots with fidelities of 81\%, 95\% and 95\%. 
We found that due to the low gate density in our atomic precision donor device, the capacitive coupling between sensor and qubits follows a $1/d^{1.4}$ dependence compared to the $1/d^{2.5-3.0}$ dependence observed in gate-defined devices that accumulate qubit electrons below metallic gates. As a consequence charge sensor range is significantly extended in our device compared to such gate-defined devices. Because of this extended range and the small size of donor qubits, we estimate that our current sensor could projectively measure \textapprox15 donor spin qubits in a linear array.
Our results highlight the prospect of large-scale quantum computing architectures with significantly reduced sensor densities in atom-scale spin qubits.

\begin{table*}[hbtp]
	\centering
	\caption{\textbf{Improvement in estimated qubits per sensor compared to previous single-lead readout results.}}
	\begin{tabular}{c|c|c|c|c|c|c|c}
		\hline
		\hline
		Reference & $\tau_{min}$ & \makecell{Sensor\\type} & \makecell{Device \\ type} & \makecell{Max spin \\ readout fidelity} & \makecell{Readout \\ basis} & \makecell{Estimated \\ capacitive scaling} & \makecell{Estimated qubits \\ per sensor} \\
		
		\hline
		\makecell{Pakkiam et. al., \\ (2018) \cite{Pakkiam2018}} & $19\mu$s & \makecell{Direct \\ dispersive} & \makecell{Precision \\ donor} & 83\% & \makecell{Singlet/\\triplet} & $1/d^{1.4}$ & 1 \\
		
		\hline
		\makecell{West et. al., \\ (2019) \cite{West2019}} & $3000\mu$s & \makecell{Direct \\ dispersive} & \makecell{Planar \\ MOS} & 73\% & \makecell{Singlet/\\triplet} & $1/d^3$ & 1 \\
		
		\hline
		\makecell{Zheng et. al., \\ (2019) \cite{Zheng2019}} & 170\,ns & \makecell{Direct \\ dispersive} & \makecell{SiGe} & 98\% & \makecell{Singlet/\\triplet} & $1/d^3$ & 1 \\
		
		\hline
		\makecell{Urdampiletta et. al., \\ (2019) \cite{Urdampilleta2019}} & $8\mu$s & \makecell{Latched\\ SLQD} & \makecell{Nanowire \\ MOS} & 98\% & \makecell{Singlet/\\triplet} & $1/d^{2.5-2.8}$ & 1-3 \\

		\hline
		\makecell{Chanrion et. al., \\ (2020) \cite{Chanrion2020}} & $21\mu$s & \makecell{SLQD} & \makecell{Nanowire \\ MOS} & N/A & N/A & $1/d^{2.5-2.8}$ & 3 \\
		
		\hline
		\makecell{Ansaloni et. al., \\ (2020) \cite{Ansaloni2020}} & $17\mu$s & \makecell{SLQD} & \makecell{Nanowire \\ MOS} & N/A & N/A & $1/d^{2.5-2.8}$ & 3 \\
		
		\hline
		\makecell{Cirano-Tejel et. al., \\ (2020) \cite{Cirianotejel2020}} & $50\mu$s & \makecell{SLQD} & \makecell{Nanowire \\ MOS} & N/A & \makecell{Single \\ spin} & $1/d^{2.5-2.8}$ & 3 \\
		
		\hline
		\makecell{Duan et. al., \\ (2020) \cite{duan2020}} & $9\mu$s & \makecell{SLQD} & \makecell{Nanowire \\ MOS} & N/A & N/A & $1/d^{2.5-2.8}$ & 3 \\
		
		\hline
		\makecell{\textbf{Hogg et. al.,} \\ \textbf{(this work)}} & \textbf{720\,ns} & \makecell{\textbf{SLQD}} & \makecell{\textbf{Precision} \\ \textbf{donor}} & \textbf{95\%} & \makecell{\textbf{Single} \\ \textbf{spin}} & $\mathbf{1/d^{1.4}}$ & \textbf{15} \\
		
		\hline
		\hline
		
	\end{tabular}%
	\label{tab:literature_comparison}%
\end{table*}


\section{Methods}
\textbf{Device fabrication}
The device is fabricated on a p-type natural Si substrate (1-10\si{\ohm}cm) using an STM to perform atomic precision hydrogen resist lithography \cite{Fuechsle2010}. A fully terminated H:Si(2$\times1$) reconstructed surface is prepared in an ultra-high vacuum chamber (\textapprox$10^{-12}$\,mbar). The STM tip is used to selectively remove hydrogen atoms to create a lithographic mask which defines the device. The regions of desorbed hydrogen are n-doped with $^{31}$P by introducing a gaseous $\mathrm{PH_3}$ precursor into the chamber, which adsorbs to the silicon surface only where the hydrogen has been removed. Subsequent annealing (at 350\si{\degreeCelsius}) incorporates the $^{31}$P atoms into the silicon crystal. Removing only a few hydrogen atoms allows the fabrication of atomic-scale donor quantum dots. The device is then encapsulated with 50\,nm of silicon, grown by molecular beam epitaxy. The resulting structure consists of the qubit device embedded in crystalline silicon. The device is then removed from the STM chamber and contacted electrically with aluminium ohmic contacts though etched via holes.

\textbf{Measurement details}
The device chip is bonded to a PCB to deliver high-frequency signals and DC voltages then mounted to the cold finger of a dilution refrigerator with \textapprox80\,mK base temperature.
We use a NbTiN superconducting spiral inductor for reflectometry measurements, with a resonance frequency of \textapprox130\,MHz and a loaded quality factor of \textapprox400 when bonded to L1. An AC signal is first attenuated before being applied to L1, with the reflected signal being separated to the output chain using a directional coupler, before amplification (at 4K using a CITLF3 low-noise amplifier, and at room temperature with a Pasternack PE15A1012), demodulation (Polyphase AD0105B) and acquisition with a digitiser card (NI-USB-6363). Spin readout pulse sequences are generated by an AWG (Tektronix AWG5204).

\section{Acknowledgements}
This research was conducted by the Australian Research Council Centre of Excellence for Quantum Computation and Communication Technology (project number CE170100012) and Silicon Quantum Computing Pty Ltd. MYS acknowledges an Australian Research Council Laureate Fellowship. The device was fabricated in part at the NSW node of the Australian National Fabrication Facility (ANFF).

\section{Author contributions}
MRH conceived the project and designed the device with input from PP. MRH fabricated the device with assistance from YC. MRH performed the measurements and data analysis with input from MGH. AVT designed and fabricated the superconducting inductor used for reflectometry. SKG calculated the spin readout fidelities. GKG assisted with the spin relaxation measurements. PP and MRH performed the COMSOL modelling. MRH, SKG and MYS wrote the manuscript with input from MGH. MGH and MYS supervised the project.

\section{Author information}
Correspondence and requests for materials should be addressed to MYS (michelle.simmons@unsw.edu.au).

\section{Competing financial interests}
MYS is a director of the company Silicon Quantum Computing Pty Ltd.
All other authors declare no competing financial interests.


%

\pagebreak
\widetext
\begin{center}
	\textbf{\large Supplemental Materials: Single-shot readout of multiple donor electron spins with a gate-based sensor}
\end{center}
\setcounter{equation}{0}
\setcounter{figure}{0}
\setcounter{table}{0}
\makeatletter
\renewcommand{\theequation}{S\arabic{equation}}
\renewcommand{\thefigure}{S\arabic{figure}}
\renewcommand{\thetable}{S\arabic{table}}

\section{1: Universal sensitivity metric}

To be able to compare the performance of various charge sensors in the literature, we employ the integration time for a detection signal to noise ratio of 1 ($\tau_{min}$) as our metric.
Assuming a white noise distribution, $\tau_{min}$ can be calculated from the amplitude signal to noise ratio between the two sensor signal levels measured for a given integration time ($\tau_{int}$):
\begin{equation}
	\tau_{min} = \frac{\tau_{int}}{\mathrm{SNR^2}},
	\label{eq:measurement_time}
\end{equation}
where SNR is the amplitude signal to noise ratio ($A/\sigma$, with $A$ is the signal amplitude between the 
two sensor readout levels and $\sigma$ one standard deviation of the noise distribution, assumed here to 
be white Gaussian). $\tau_{min}$ can thus easily be estimated based on standard experimental parameters. $\tau_{min}$ can be measured more accurately (at the cost of additional experimental effort) by measuring the sensor SNR for a range of $\tau_{int}$ and extrapolating to SNR=1.

We note that a charge sensitivity metric that has been commonly used in the semiconductor qubit community is electrons per root Hertz ($e/\mathrm{\sqrt{Hz}}$) \cite{Gonzalez-Zalba2015, Ahmed2018, Schaal2020, Stehlik2015, Zheng2019}. However there are two distinct methods for calculating the charge sensitivity in $e/\mathrm{\sqrt{Hz}}$ which in general do not give equivalent results and make direct comparison between experiments difficult. The first is a frequency-domain method, where a small modulating signal is applied to a gate nearby the sensor, and the magnitude of frequency sidebands generated by the modulation signal is measured \cite{Schoelkopf1998}. This frequency-domain method for determining the charge sensitivity in $e/\mathrm{\sqrt{Hz}}$ is scaled to the charging energy of the sensor (which can vary significantly between devices and different material systems) and hence cannot be directly compared between different experiments. This method also only quantifies the response of the sensor to small charge shifts (much smaller than the Coulomb peak linewidth), and is hence an inappropriate metric for strong-response regime charge sensors (where the sensor signal shifts by an appreciable fraction of a linewidth due to a charging event on a nearby qubit) \cite{Keith2019, Chanrion2020, Ansaloni2020, Cirianotejel2020, Borjans2021}. Because it is scaled to the sensor charging energy, this frequency domain method is also unsuitable for sensors which do not have a periodic response in the gate charge, such as direct dispersive sensors measuring inter-dot transitions.

The second method for calculating the charge sensitivity in $e/\mathrm{\sqrt{Hz}}$ is to first determine $\tau_{min}$, then calculate the charge sensitivity as \cite{Stehlik2015, Zheng2019}
\begin{equation}
	S = e\sqrt{\tau_{min}}.
	\label{eq:time_domain_sensitivity}
\end{equation}
Typically when using Eq. \ref{eq:time_domain_sensitivity}, $e$ is set to 1, as the charge state of the readout target is changing by a single electron. The results from this second method are however not consistent with the frequency domain method, for which $e=1$ indicates a shift in the sensor peak of the full sensor charging energy. As a consequence the second method thus generally gives significantly larger (i.e. worse) values of $e/\mathrm{\sqrt{Hz}}$, and direct comparisons between $e/\mathrm{\sqrt{Hz}}$ values determined using the two different methods must be treated with caution.

For these reasons we thus explicitly use $\tau_{min}$ in the main text as a more useful metric than $e/\mathrm{\sqrt{Hz}}$ for directly comparing readout performance in semiconductor spin qubits. Not only is it applicable in both weak and strong-response regimes, but also for sensors that do not create a signal periodic in the gate charge such as dispersive senors. 
Table \ref{tab:taum_comparison} therefore compares $\tau_{min}$ (calculated using Eq. \ref{eq:measurement_time} where not directly reported) for a selection of recent dispersive readout results. 
For reference, the current best charge sensor performance is Keith \emph{et. al.} \cite{Keith2019}, who achieved $\tau_{min}=0.62$\,ns (SNR=12.7, $\tau_{int}=100$\,ns) with an rf-SET.

\begin{table}[hbtp]
	\centering
	\caption{\textbf{Sensitivity comparison of reported single-lead sensing results}}
	\setlength{\tabcolsep}{0.3em} 
	{\renewcommand{\arraystretch}{1.5}
		\begin{tabular}{l|c|c|c|c|c|c|c}
			\hline
			\hline
			
			Reference & $\tau_{int}$ & SNR & $\tau_{min}$ & \makecell{Sensor \\ type} & \makecell{Resonator \\ type} & \makecell{Resonator \\ frequency} & \makecell{Lever\\arm} \\
			
			\hline
			\makecell{\textbf{Hogg et. al.,} \\ \textbf{2020 (this work)}} & \textbf{66\,$\mu$s} & \textbf{9.6} & \textbf{720\,ns} 	& \textbf{SLQD}  		& 
			\makecell{\textbf{NbTiN} \\ \textbf{spiral}}		 & \textbf{130\,MHz} & \makecell{\textbf{0.38$\pm$0.05} \\ (1-$\alpha_{_{\mathrm{L1}}}$) }\\
			
			\hline
			\makecell{Pakkiam et. al., \\ 2018 \cite{Pakkiam2018}} & 300\,$\mu$s & 4 & 19$\mu$s  	& \makecell{Direct\\dispersive} & \makecell{NbTiN \\ spiral}  & 335\,MHz & \makecell{0.13$\pm$0.05 \\ ($\Delta\alpha$)} \\
			
			\hline
			\makecell{West et. al., \\ 2019 \cite{West2019}} & 12\,ms & 2 & 3000$\mu$s & \makecell{Direct\\dispersive} & \makecell{PCB \\ inductor} & 267\,MHz & \makecell{0.1$\pm$0.03 \\ ($\Delta\alpha$) } \\
			
			\hline
			\makecell{Urdampellita \\ et. al., 2019 \cite{Urdampilleta2019}} & 500\,$\mu$s & 6.5 & 8$\mu$s  & \makecell{SLQD\\(latched)} & \makecell{PCB \\ inductor} & 234\,MHz & \makecell{0.3 \\($\alpha$)} \\
			
			\hline
			\makecell{Zheng et. al., \\ 2019 \cite{Zheng2019}} & 170\,ns & 1 & 	170\,ns	& \makecell{Direct\\dispersive}  & \makecell{NbTiN\\on-chip}  & 5.7\,GHz & - \\

			\hline
			\makecell{Stehlik et al., \\ 2015 \cite{Stehlik2015}} & 7\,ns & 1 & 	7\,ns  & \makecell{Direct\\dispersive (JPA)} & \makecell{Nb \\on-chip}	 & 7.88\,GHz & - \\
			
			\hline
			\makecell{Schaal et al., \\ 2020 \cite{Schaal2020}} & 80\,ns & 1 & 	80\,ns & \makecell{Direct\\dispersive (JPA)} & \makecell{NbN \\ spiral}  & 622\,MHz & \makecell{0.36 \\ ($\Delta\alpha$)} \\

			\hline
			\hline
		\end{tabular}%
	}
	\label{tab:taum_comparison}%
\end{table}

The lever arm column in Table \ref{tab:taum_comparison} lists the relevant lever arm factor for the dispersive sensor, which depends on the type of sensor. For example, for direct dispersive sensors the amount of signal generated depends on the \emph{differential} lever arm ($\Delta\alpha$) between the gate to which the probe rf signal is applied and the two quantum dots. For SLQD sensors the amount of signal generated depends on the lever arm of the reflectometry gate acting on the sensor dot ($\alpha$, or $1-\alpha$ if the reflectometry gate is also the reservoir for the sensor dot).
Table \ref{tab:taum_comparison} highlights that dispersive sensor performance can be optimised by increasing the resonator frequency, using a low-noise amplifier (such as a Josephson parametric amplifier, JPA) and maximising the relevant lever arm factor.

\FloatBarrier

\section{2. Optimising the sensor for single-electron charge detection}
\begin{figure}[!hbtp]
	\includegraphics[width=0.5\textwidth]{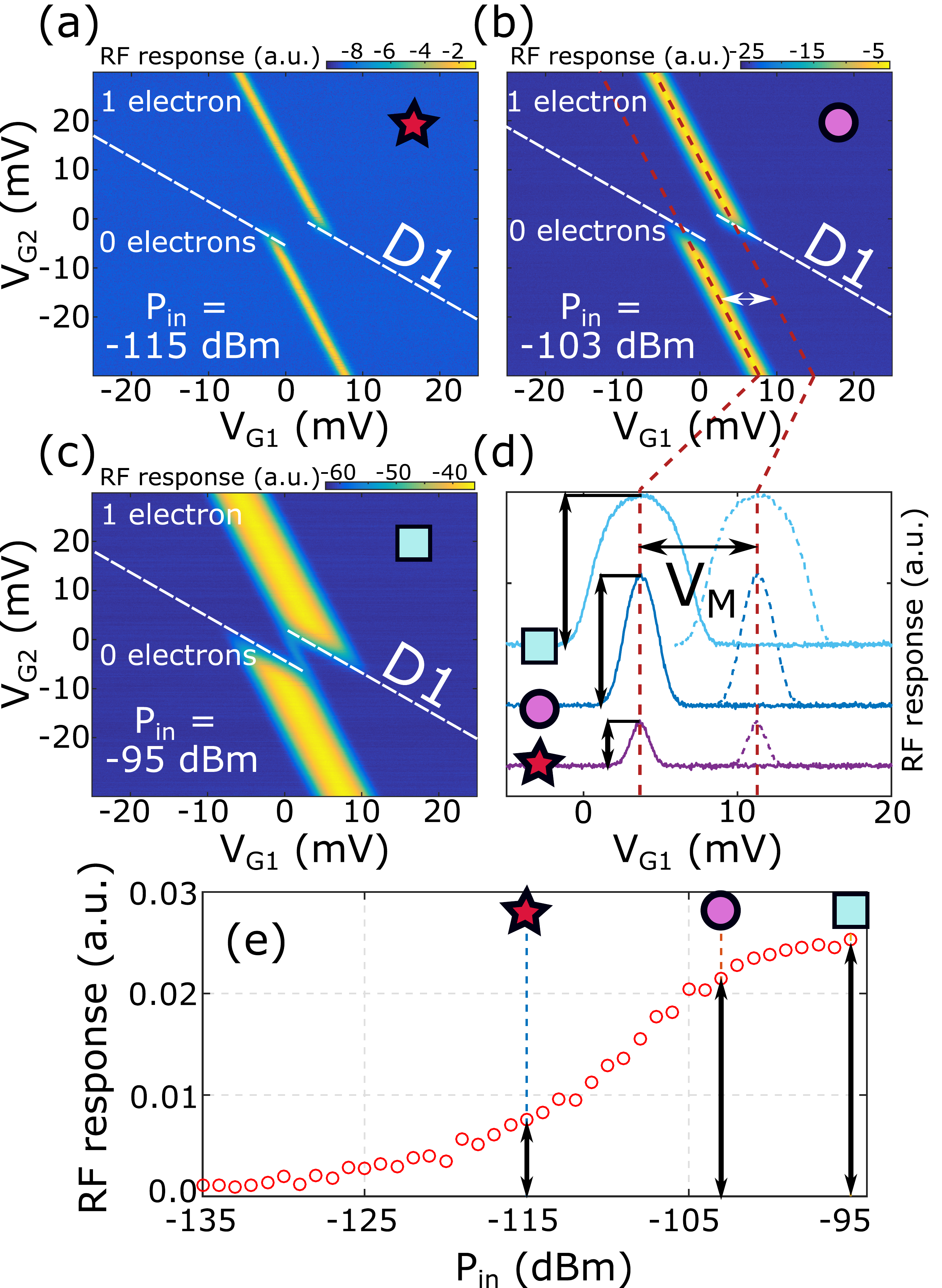}%
	\caption{\textbf{Optimising drive power of SLQD sensor for single-electron detection.} Gate-gate maps with $V_\mathrm{G1}$ and $V_\mathrm{G2}$ around the first electron charge transition of D1 for multiple $P_\mathrm{in}$ values. Each map shows a clear break in the SLQD1 transition due to D1 charging. $P_\mathrm{in}$ values are \textbf{(a)} -115\,dBm, \textbf{(b)} -103\,dBm and \textbf{(c)} -95\,dBm, for which power broadening of the Coulomb peak is apparent. The power value in (b) gives an optimal balance between power broadening and signal amplitude.
		\textbf{(d)} SLQD1 transition line for 0 and 1 electron D1 charge states for the three power levels in (a)-(c). 
		Blue square matches the power from (c), purple circle from (b) and red star from (a). $V_\mathrm{M}$ indicates the shift in SLQD sensor response due to D1 charging. For this transition $V_\mathrm{M}=7.1$\,mV. 
		\textbf{(e)} Peak SLQD1 signal amplitude as a function of input power, with the power levels from (a),\,(b) and (c) indicated with the same shapes as (d). Saturation is observed at high powers, $P_\mathrm{in} > -103$\,dBm.}
	\label{fig:power_optimisation_figure}
\end{figure}

To optimise SLQD1 for the time-resolved charge detection of the donor quantum dots, the main experimental parameter we can tune is the input reflectometry power $P_\mathrm{in}$. In an SLQD the sensor signal saturates as a function of $P_\mathrm{in}$ \cite{House2016}. 
This can be understood intuitively by considering the cyclical single-electron tunnelling process that generates the dispersive signal. When $P_\mathrm{in}$ is large enough to fully traverse a Coulomb peak, a full electron is driven between L1 and SLQD1 each time the reflectometry signal reverses polarity. Because there is no DC current path, the tunnelling current is limited to two electrons per AC cycle.
The amplitude of the tunnelling current is thus constrained by Coulomb blockade. The measured signal is directly proportional to the tunnelling current in the device, hence this also saturates. In our measurements choosing $P_\mathrm{in}$ to be at the onset of this signal saturation was optimal for readout (purple circle in Fig. \ref{fig:power_optimisation_figure}\,(d)), as we now discuss.

Figures \ref{fig:power_optimisation_figure}\,(a)-(c) show identical gate-gate maps around the first electron charge transition of D1 for three different $P_\mathrm{in}$ levels. 
Figure \ref{fig:power_optimisation_figure}\,(a) has $P_\mathrm{in}$ below the saturation level,\,(b) is at the onset of saturation and (c) is well into saturation. 
Beyond the saturation point increasing the input power does not return significantly more signal, and the Coulomb peaks start to power broaden.
To sense our donor electrons, we choose the $P_\mathrm{in}$ value from \ref{fig:power_optimisation_figure}\,(b), which gives an optimal balance between signal contrast and Coulomb peak power broadening.

Figure \ref{fig:power_optimisation_figure}\,(d) demonstrates the shift in the SLQD1 sensor response ($V_\mathrm{M}$) due to an electron charging event on D1. The size of this shift depends on the capacitive coupling between SLQD1 and the target qubit, and for D1 is $V_\mathrm{M}=7.1$\,mV. As discussed in the main text, the size of $V_\mathrm{M}$ as a function of distance to the sensor is an important factor in determining the density of sensors required in a device. An architecture in which $V_\mathrm{M}$ decreases slowly as a function of distance can reduce the number of charge sensors required. 
We note that due to the strong capacitive coupling between SLQD1 and D1, $V_\mathrm{M}$ is larger than a Coulomb peak width for all $P_\mathrm{in}$ values in Fig. \ref{fig:power_optimisation_figure}.
In this situation the maximum signal contrast for sensing D1 is thus obtained by tuning to the top of a Coulomb peak, rather than the side of a peak which would give the best small-signal sensitivity. This is called the strong-response charge sensing regime \cite{Keith2019}, and allows binary on-off switching of the full sensor signal during detection. 

In our experiment all four donor quantum dots are in the strong response regime for $P_\mathrm{in}=-103$\,dBm as shown in \ref{fig:power_optimisation_figure}\,(b), hence we use the same $P_\mathrm{in}$ to sense all four qubits. 
Figure \ref{fig:power_optimisation_figure}\,(e) shows the maximum sensor contrast as a function of $P_\mathrm{in}$, with the values from (a), (b) and (c) indicated by the matching coloured shapes. For $P_\mathrm{in}>-103$\,dBm the maximum signal saturates, and going above this value causes power broadening without any significant gain in signal. This further justifies our use of $P_\mathrm{in}=-103$\,dBm.

\FloatBarrier

\section{3. Readout fidelity calculation}

In this section we detail the method used to estimate the measurement fidelity in our experiments using the theory outlined in Keith \emph{et. al.} \cite{Keith2019a}. For completeness we summarise the fidelity calculations below.

The measurement fidelity can be broken down into two different processes in the detection event. The first process is the probability that the $\ket{\uparrow}$ electron tunnels to the reservoir and not the $\ket{\downarrow}$. This stage is known as state-to-charge conversion (STC) and is used to optimise the measurement time, $t_{opt}$. The second stage relates to the probability of correctly measuring a tunneling event, known as electrical detection (E) and sets our optimal threshold voltage, $V_{opt}$. From these two processes, four fidelities can be obtained for each spin state/process combination, $F^{\downarrow}_{STC}$,  $F^{\uparrow}_{STC}$, $F^{\downarrow}_{E}$, and $F^{\uparrow}_{E}$. Using these fidelities we can calculate the individual state measurement fidelities using,

\begin{align}
	F_{\downarrow} &= F^{\downarrow}_{STC} F^{\downarrow}_{E} + (1 - F^{\downarrow}_{STC})(1 - F^{\uparrow}_{E}),\\
	F_{\uparrow} &= F^{\uparrow}_{STC} F^{\uparrow}_{E} + (1 - F^{\uparrow}_{STC})(1 - F^{\downarrow}_{E}).
\end{align}
Finally, the total measurement fidelity is the average between the two individual state fidelities,

\begin{equation}
	F_M = \frac{F_{\downarrow} + F_{\uparrow}}{2}.
\end{equation}

To calculate the measurement fidelity the following parameters were obtained from the experiment: $\mu_0$ and $\mu_1$ are the low and high signal levels of the SLQD respectively with corresponding noise power spectral densities (assuming white Gaussian noise) $A_0$ and $A_1$. $f_c$ is the filter cut-off frequency of the sensor, $\Gamma_s$ is the digitising sample rate of the data acquisition hardware used in the experiment, $t^{\alpha}_{\textnormal{IN/OUT}}$ are the tunnel times of the electrons in and out of the electron reservoir ($\alpha = \downarrow, \uparrow$), $T_1$ is the relaxation time of the $\ket{\uparrow}$ state and $B_z$ is the external magnetic field. In Table~\ref{tab:LandRpar} we state all the parameters and metrics used to calculate the spin readout fidelity for each quantum dot measured in the main text.

\begin{table}[htbp!]
	\caption{\centering\textbf{Parameters used in the determination of the single-shot readout fidelity for qubits in the main text.}}
	\begin{tabular}{c | c | c | c}
		\hline
		\hline
		Parameter			 		&  D1 value & D2 value & D3 value\\
		\hline
		$\mu_0$ 		 			& -0.0453$\pm$0.0001 & -0.0371$\pm$0.0001 & -0.0986$\pm$0.0001\\
		$\mu_1$			 			& -0.0302$\pm$0.0001 & -0.0142$\pm$0.0001 & -0.0623$\pm$0.0001\\
		$A_0$ 						& (1.02$\pm$0.01) $\times 10^{-5}$ $1/\sqrt{\textnormal{Hz}}$ & (2.67$\pm$0.01) $\times 10^{-5}$ $1/\sqrt{\textnormal{Hz}}$ & (3.24$\pm$0.01) $\times 10^{-5}$ $1/\sqrt{\textnormal{Hz}}$\\
		$A_1$ 						& (1.08$\pm$0.01) $\times 10^{-5}$ $1/\sqrt{\textnormal{Hz}}$ & (2.72$\pm$0.04) $\times 10^{-5}$ $1/\sqrt{\textnormal{Hz}}$ & (3.63$\pm$0.03) $\times 10^{-5}$ $1/\sqrt{\textnormal{Hz}}$\\
		$f_c$ 						& 80 kHz & 15 kHz & 15 kHz\\
		$\Gamma_s$ 					& 500 kHz & 50 kHz & 50 kHz\\
		$t^{\uparrow}_{\textnormal{OUT}}$ 	& 65$\pm$9 $\mu$s & 499$\pm$4 $\mu$s & 744$\pm$12 $\mu$s\\
		$t^{\downarrow}_{\textnormal{OUT}}$ 	& 7.9$\pm$2.3 ms & 254$\pm$26 ms & 66$\pm$5 ms\\
		$t^{\downarrow}_{\textnormal{IN}}$ 	& 15$\pm$0.5 $\mu$s & 257$\pm$6 $\mu$s & 293$\pm$9 $\mu$s\\
		$T_1$						& 1.5$\pm$0.4 s & 38$\pm$3 ms & 11.6$\pm$4.0 s\\
		$B_z$ 						& $1.5$ T & $1.5$ T & $1.5$ T\\
		\hline
		$t_{opt}$ (ms)			 	& 0.31$\pm$0.01 & 3.1$\pm$0.1 & 3.4$\pm$0.1\\
		V$_{opt}$ (arb.)		 	& -0.0322$\pm$0.0001 & -0.0196$\pm$0.0001 & -0.0759$\pm$0.0003\\
		$F_{STC}^{\downarrow}$ (\%)			& 96.1$\pm$1.4 & 98.8$\pm$0.2 & 95.0$\pm$0.4\\
		$F_{STC}^{\uparrow}$ (\%)			& 99.2$\pm$0.5 & 98.6$\pm$0.2 & 98.9$\pm$0.1\\
		$F_{E}^{\downarrow}$ (\%)				& 95.3$\pm$0.3 & 99.1$\pm$0.1 & 99.7$\pm$0.1\\
		$F_{E}^{\uparrow}$ (\%)				& 69.1$\pm$0.9 & 92.9$\pm$0.1 & 95.3$\pm$0.1\\
		$F_{\downarrow}$ (\%)					& 92.8$\pm$0.5 & 98.0$\pm$0.1 & 94.9$\pm$0.3\\
		$F_{\uparrow}$ (\%)					& 68.6$\pm$1.2 & 91.6$\pm$0.3 & 94.3$\pm$0.2\\
		$F_M$ (\%)					& 80.7$\pm$1.1 & 94.8$\pm$0.2 & 94.6$\pm$0.3\\
		\hline
		\hline
	\end{tabular}
	\label{tab:LandRpar}
\end{table}

\section{4. COMSOL simulation of patterned device}
\label{sec:simulation_description}

In Fig. 3\,(a) of the main text we presented simulation results showing $V_\mathrm{M}$ as a function of spatial position for the device in Fig. 1\,(a) of the main text. Here we provide details of the simulation, which was performed using the finite element solver COMSOL multiphysics.

The shift in the SLQD sensor Coulomb peak due to a single-electron charging event on a target qubit is determined by the mutual charging energy ($E_\mathrm{M}$) between the sensor and qubit. To find $E_\mathrm{M}$ we placed a charge of one electron on SLQD1 and simulated the resulting change in potential throughout the device. Here we note that the change in potential at a given spatial position $(x_1, y_1)$ when adding an electron to SLQD1 is equivalent to the change in potential at SLQD1 when adding an electron to a qubit located at $(x_1, y_1)$. Our simulation of adding an electron to SLQD1 thus directly returns us a spatial map of the mutual charging energy throughout the device. To convert the simulated $E_\mathrm{M}$ into $V_\mathrm{M}$ (the mutual charging voltage in terms of $V_\mathrm{G1}$) for direct comparison with the experimental $V_\mathrm{M}$ shifts from Figs. 2\,(a),(c),(e),(g) in the main text, we divided $E_\mathrm{M}$ by the measured G1$\rightarrow$SLQD1 lever arm $\alpha_{G1}^{SLQD1}=0.09\pm0.01$. We then accounted for an observed offset between the simulation-derived $V_\mathrm{M}$ results and the experimental data by calibrating the simulation results to match the experimental data at the position of D1. After this point calibration, the simulation data matched the experimental data at the positions of D2, D3 and D4, giving us confidence in the simulation method.

\begin{figure}[!hbtp]
	\includegraphics[width=0.55\textwidth]{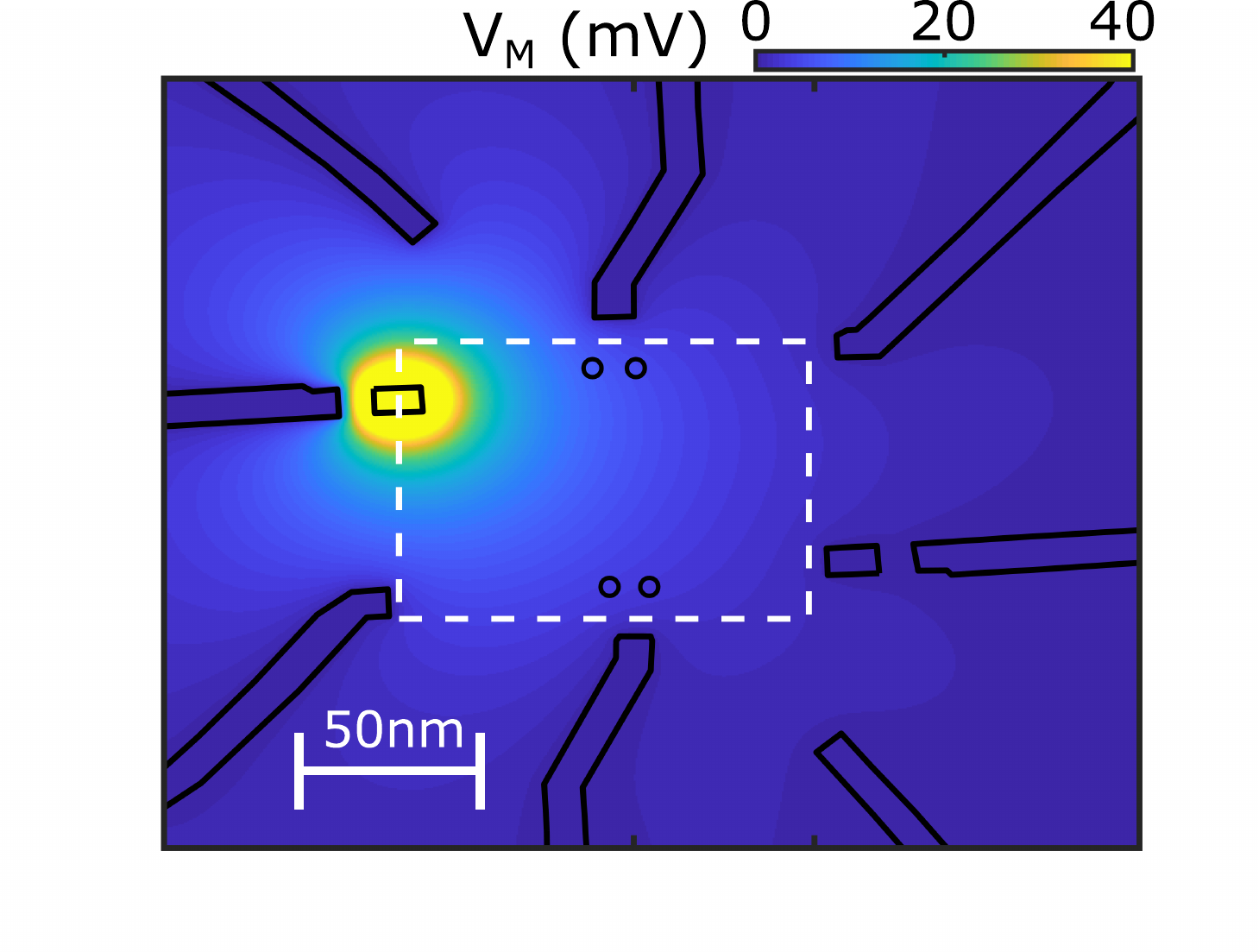}%
	\caption{\textbf{Active device region used to extract distance scaling.} Inside the white dashed box we calculated the distance from each simulation point to the center of SLQD1, and used the corresponding $V_\mathrm{M}$ values to generate Fig. 3\,(b) in the main text.
	}
	\label{fig:box_region}
\end{figure}

To obtain the plot of $V_\mathrm{M}$ as a function of distance to SLQD1 shown in Fig. 3\,(b) of the main text, we used the $V_\mathrm{M}$ values inside the region bounded by the white dashed box in Fig. \ref{fig:box_region}. We chose the device region inside the white dashed box to include the active sensing region, but avoid including regions where no donor would be patterned in a real device (e.g. directly behind gates or within \textapprox10\,nm of a gate).
For each simulation point inside the white dashed box in Fig. \ref{fig:box_region}, we calculated the distance from the center of SLQD1. Inside the white dashed box, we found that $V_\mathrm{M}$ followed a $1/d^{1.4\pm0.1}$ dependence as a function of distance to SLQD1.

\section{5. Long-range readout fidelity and distant qubit sensing}

To calculate the fidelity as a function of the qubit-sensor distance, we use the expected $V_\mathrm{M}$ shift in sensor response for a given $d$ based on the $1/d^{1.4}$ fit in Fig. 4\,(b). The maximum difference between the two peaks separated by $V_\mathrm{M}$ then gives $\Delta V(d)$, the maximum sensor contrast as a function of $d$. To calculate the fidelity we modify the $\mu_1$ level such that,
\begin{equation}
	\mu_1(d) = \mu_0 + \mu_1(0) \frac{\Delta V(d)}{\Delta V(0)},
\end{equation}
where $\Delta V(0)$ is the full peak height (i.e. signal contrast when $V_\mathrm{M}$ is large enough that the peaks are fully separated). In determining $\Delta V(d)$, we use the Coulomb peak shape extracted from fitting 
our experimental Coulomb peaks. In this way we calculated the sensor signal contrast as a function of $d$, the qubit-sensor distance. We then use $\mu_1(d)$ in our fidelity calculation with the experimental spin-dependent tunnel rates and $T_1$ relaxation time measured for D3, which are unrelated to the qubit-sensor distance.

Table II in the main text compares single-lead sensor results from the literature, including the device type and an estimate of the number of qubits that could be read by one sensor for the specified sensor type and fabrication platform. In compiling this table we use the reported value for $\tau_{min}$ (or use Eq. \ref{eq:measurement_time} where $\tau_{min}$ is not explicitly stated). For the planar gate-defined devices we use the $1/d^3$ capacitive scaling that has been previously reported \cite{Zajac2016, Neyens2019}, and for nanowire MOS devices we use the estimate of $1/d^{2.5-2.8}$ based on the device modelling from \cite{duan2020}. We note that in some of the results of ref. \cite{duan2020} the capacitive coupling as a function of distance was increased using metallic couplers, which we neglect in the present analysis to focus on the bare sensor range. In estimating the number of qubits per sensor we use published literature values for geometric parameters (e.g. 80\,nm qubit pitch for nanowire MOS \cite{Chanrion2020}). For the results in which no spin readout was performed, we assume qubit parameters identical to D3 in our device and a sensor identical to SLQD1. We then estimate the number of qubits that could be read out with $>90\%$ fidelity based on the geometry and appropriate capacitive scaling. We assume a 2xN array of qubits for the nanowire MOS devices (as such arrays naturally form in split-gate geometries \cite{Chanrion2020}), and a 1xN array for all others device types.

\section{6. COMSOL simulation of linear array of donor qubits}

\begin{figure}[!tb]
	\includegraphics[width=0.9\textwidth]{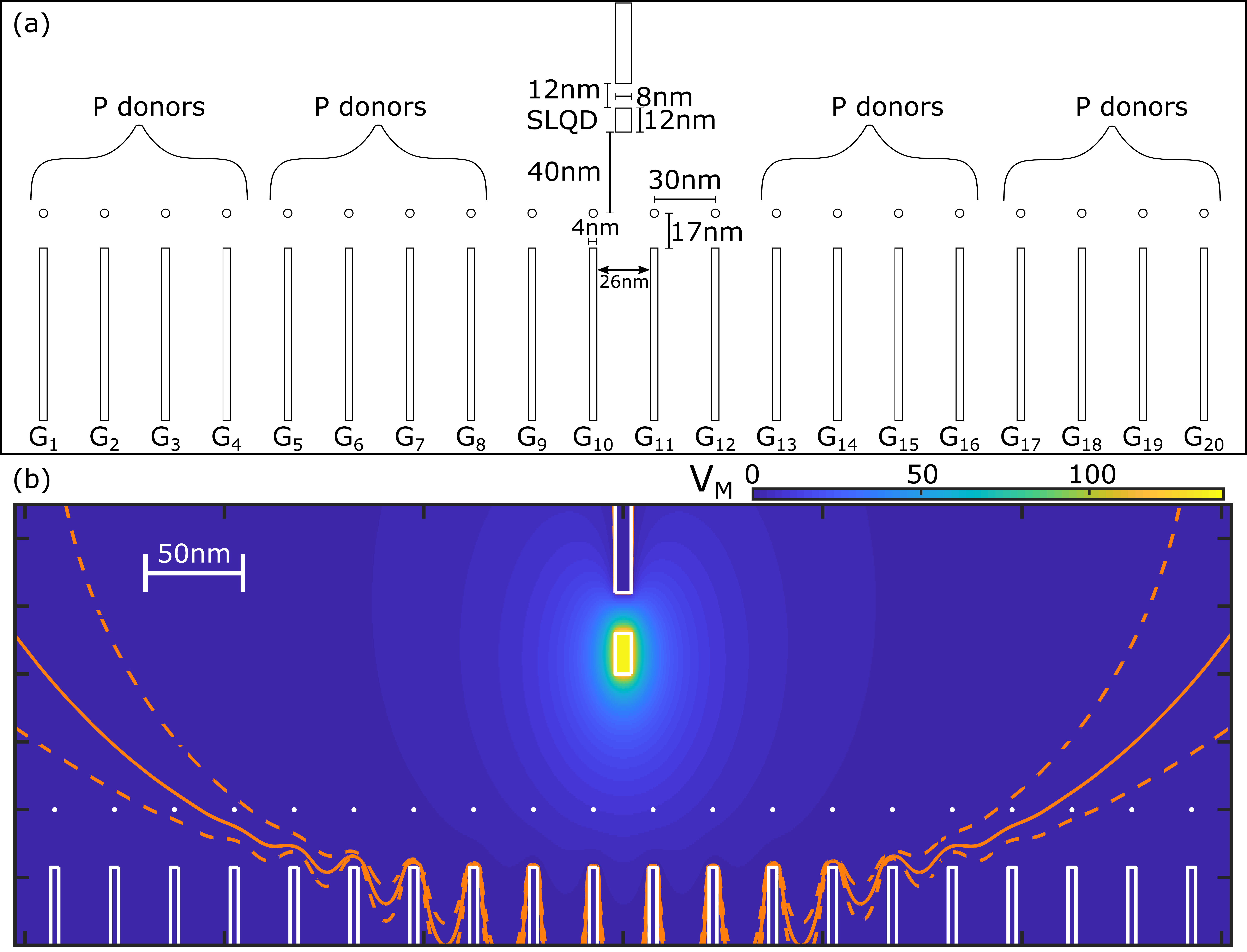}%
	\caption{\textbf{COMSOL simulation of $V_\mathrm{M}$ across a 20-dot linear array.}
		\textbf{(a)} Schematic of a linear array geometry simulated using COMSOL. The array contains 20 donor quantum dots (with a 30\,nm spacing between donors), with gates G1-G20 controlling the donor potentials and serving as electron reservoirs. A single SLQD in the center of the array operates as a charge sensor.
		\textbf{(b)} Electrostatic model of $\mathrm{V_M}$ for the architecture shown in (a). The orange contour line shows the boundary of 90\% readout fidelity using an SLQD with the same performance as that demonstrated in the main text. The dashed lines show the uncertainty bounds of the 90\% fidelity contour. We find that 14-16 donors are contained within the uncertainty region, indicating that a single SLQD sensor could read \textapprox15 donor qubits in this linear array geometry.
	}
	\label{fig:50dot_simulation}
\end{figure}

In the main text we estimated that based on a simple extrapolation of the 1/$d^{1.4}$ sensor capacitive scaling that we observed in our experiment, it should be feasible to read out $\lesssim$20 donor qubits in a linear array using a single sensor. Here we present COMSOL modelling of a linear qubit array to improve the accuracy of this estimate using a simulated device with realistic dimensions. Figure \ref{fig:50dot_simulation}\,(a) shows a schematic of the linear array architecture. The array contains 20 P donor quantum dots, with an inter-dot separation of 30\,nm (we chose 30\,nm as this is the separation used in a leading donor surface code proposal \cite{Hill2015}). In ref. \cite{Hill2015} qubits are encoded in the nuclear spin states of single P donors, with single-qubit gates applied using global NMR and two-qubit gates mediated by an electron magnetic dipole interaction (see ref. \cite{Hill2015} for full details). For the array shown in Fig. \ref{fig:50dot_simulation}, qubits could also potentially be encoded in electron spin states, with single-qubit gates achieved via global ESR (addressed via electric field tuning the Stark shift \cite{Rahman2007, Laucht2015}) and magnetic dipole coupling for two-qubit gates.

Gates G1-G20 provide electrostatic tuning of the donor potentials, as well as serving as electron reservoirs. Each gate serves as the reservoir for the nearest donor. A SLQD is positioned in the center of the array to sense the qubits. Figure \ref{fig:50dot_simulation}\,(b) shows an electrostatic simulation of $\mathrm{V_M}$ (calculated using the same method described in Section S4) for the linear array architecture in Fig. \ref{fig:50dot_simulation}\,(a). 

Here the orange contour line shows the boundary for which a qubit (with the same tunnel rates and T$_1$ time as D3) could be read with over 90\% fidelity using a sensor with the same performance as SLQD1. The dashed lines show the uncertainty bounds of the 90\% fidelity contour.
In Fig. \ref{fig:50dot_simulation}\,(b), 14-16 donors are contained within the uncertainty bounds. This direct COMSOL modelling of a donor linear array thus indicates that 14-16 donor qubits could be read out using an SLQD sensor with the same performance as our experimental sensor, slightly fewer than the $\textapprox$20 estimated from extrapolating the $1/d^{1.4}$ fit in Fig. 3\,(b) in the main text. We note that the linear array geometry shown in Fig. \ref{fig:50dot_simulation}\,(a) is fully consistent with previously demonstrated lithographic feature sizes using atomic precision fabrication techniques \cite{Weber2012}.

\end{document}